\documentclass[preprint,aps,amsmath,amssymb]{revtex4}
\usepackage{graphicx}
\usepackage{dcolumn}
\usepackage{bm}
\begin{document}


\title{Enhancing capacity of coherent optical information storage and transfer in a Bose-Einstein condensate}

\author{\"O. E. M\"ustecapl{\i}o\u{g}lu$^{1}$}
\email{omustecap@ku.edu.tr,dtarhan@gmail.com}
\author{Devrim Tarhan$^{1,2,3}$}
\affiliation{$^{1}$Department of Physics, Ko\c{c} University,
Sar{\i}yer, Istanbul 34450, Turkey\\$^2$ Department
of Physics, Istanbul Technical University, Maslak, 34469, Istanbul, Turkey \\
$^3$ Department of Physics, Harran University, \c{S}anl\i{}urfa,
Turkey}
\begin{abstract}
Coherent optical information storage capacity of an atomic
Bose-Einstein condensate is examined. Theory of slow
light propagation in atomic clouds is generalized to short pulse
regime by taking into account group velocity dispersion. It is
shown that the number of stored pulses in the condensate can be
optimized for a particular coupling laser power, temperature and
interatomic interaction strength. Analytical results are derived
for semi-ideal model of the condensate using effective uniform
density zone approximation. Detailed numerical simulations are
also performed. It is found that axial density
profile of the condensate protects the pulse against the group velocity dispersion.
Furthermore, taking into account finite radial size of the
condensate, multi-mode light propagation in atomic Bose-Einstein condensate is investigated.
The number of modes that can be supported by a condensate is
found. Single mode condition is determined as a function of
experimentally accessible parameters including trap size,
temperature, condensate number density and scattering length.
Quantum coherent atom-light interaction schemes are proposed for
enhancing multi-mode light propagation effects. \\
\\
Topic: Physics of Cold Trapped Atoms (Report number: 6.2.4)
\end{abstract}
\maketitle
\section{Introduction}

Light can be slowed down via electromagnetically induced
transparency (EIT) in an atomic medium. This effect is solely
due to the steep dispersion at the EIT resonance. It has been demonstrated
both for the ultracold Bose-Einstein condensate (BEC)\cite{hau99,inouye00}
as well as for an atomic cloud at the room temperature\cite{kash99}.
In a typical experimental set up for EIT, a probe pulse is sent to an atomic
cloud, consisting of effectively three level atoms. The atomic
medium would be normally opaque to the transmission of the probe
pulse at resonance. Another, and more stronger laser pulse is used to
establish quantum coherence in the lower level doublet so that the
system is put in a dark state, in which destructive quantum
interference of probability amplitudes for upper level transitions
cancel eachother. This allows for propagation of the probe pulse in the
medium at group speeds determined by the steepness of the
dispersion curve at resonance. In the case of EIT, the dispersion
is so steep that the probe pulse can only propagate at very low
speeds in the order of few m/s.

In the experiments, group velocity is measured operationally and
indirectly from the delay time. Probe pulse is divided into two
and one pulse is propagated in free space without entering into
the atomic medium. Using this pulse as a reference, delay time for
the probe pulse is determined. The length of the atomic cloud is measured
by using an imaging laser. The ratio of the length of the medium
to the delay time, which gives the group velocity of the probe
pulse, is calculated at various temperatures. The temperature
dependence of the group velocity exhibits a sharp drop at the
critical temperature of condensation, and can be understood from
the temperature dependence of the condensate density. In order to
estimate the group velocity, one can assume uniform density over
the width of the condensate, which would be taken as the effective
length of the condensate. From the EIT susceptibility,
\begin{eqnarray}
\chi \left( {\vec r} \right) = \rho \left( {\vec r}
\right)\frac{{\left| {\mu _{31} } \right|^2 }}{{\varepsilon _0
\hbar }}\frac{{i\left( {i\Delta  + {{\Gamma _2 } \mathord{\left/
 {\vphantom {{\Gamma _2 } 2}} \right.
 \kern-\nulldelimiterspace} 2}} \right)}}{{\left[ {\left( {i\Delta  + {{\Gamma _2 } \mathord{\left/
 {\vphantom {{\Gamma _2 } 2}} \right.
 \kern-\nulldelimiterspace} 2}} \right)\left( {i\Delta  + {{\Gamma _3 } \mathord{\left/
 {\vphantom {{\Gamma _3 } 2}} \right.
 \kern-\nulldelimiterspace} 2}} \right) + {{\Omega _c^2 } \mathord{\left/
 {\vphantom {{\Omega _c^2 } 4}} \right.
 \kern-\nulldelimiterspace} 4}} \right]}},
\end{eqnarray}
one can calculate the group velocity, defined by
\begin{eqnarray}
\frac{1}{{v_g }} = \frac{1}{c} - \frac{\pi }{\lambda
}\frac{{\partial \chi }}{{\partial \omega }}\left( {\omega _0 }
\right),
\end{eqnarray}
which leads to
\begin{eqnarray}
v_g  = \frac{{c\varepsilon _0 \hbar \Omega _c^2 }}{{2\omega _0
\left| {\mu _{31} } \right|^2 \rho }}.
\end{eqnarray}
It should be noted that this is valid for pump laser powers in the
range
\begin{eqnarray}
\sqrt {\frac{{4\pi c\left| {\mu _{31} } \right|^2 \rho
}}{{\varepsilon _0 \hbar }}}  \gg \Omega _c  \gg \Gamma _{2,3}.
\end{eqnarray}

Slow light allows for novel nonlinear optical regimes and
applications. Early proposals include enhanced nonlinear optical
effects\cite{kash99}, nonlinear magneto-optics\cite{budker99},
nonlinear optics with single photons\cite{lukin00,lukin01}.
Besides, slow light is thought to be useful for optical data
storage, nonlinear optics with few photons, quantum entanglement of
slow photons, and enhancing acousto-optical effects\cite{matsko00}. These effects
may be useful in all-optical computers, telecommunication systems, and memory chips.
Most promising application of slow light is believed to be optical data storage,
whose possibility is experimentally demonstrated\cite{liu01}. While
slow light in atomic systems, particularly in BECs,
could be useful in quantum information technology, for practical quantum information processing
it is necessary to be able to inject more pulses into the atomic system
during the storage time\cite{liu01}.

A significant parameter to characterize the practical value of an
information storage device is the bit storage capacity, which
should be as high as possible for a useful dynamic memory
device. Bit storage capacity can be defined as the number of bits
that can be simultaneously present in the device during the
storage time. In the case of EIT based slow light in atomic
systems, bits correspond to optical probe pulses. In the present
experiments, both the storage time and probe pulse
widths are about few microseconds. Therefore, only single pulse can be
present in the condensate during the storage time.

It should be noted that multiple pulses of different polarization could be injected
into the atomic cloud under EIT conditions at the cost of tolerable absorption as well as
increased group speeds\cite{agarwal02}. Modification of the EIT transparency window
may also be exploited to inject multiple pulses\cite{wei05}.
In order to enhance the bit storage capacity, recent experiments and proposals indicate that
perhaps the most promising and straightforward direction could be just to use shorter
pulses\cite{sautenkov05,sun05,bashkansky05}.

One serious obstacle that hinder immediate application of short pulses into
atomic gases to enhance their information storage and transfer
capacities is that short pulses may suffer strong pulse shape
distortions due to high order dispersion effects. In particular,
second order dispersion yields pulse spread due to group velocity
dispersion. Different frequency components of the pulse travel at
different speeds resulting in broadening of the pulse over time.
Using nanosecond pulses, for example, instead of microsecond
ones, may not necessarily increase the bit storage capacity $1000$
times more. To see this quantitatively, let us consider an ideal pulse
repetition rate as ${1 \mathord{\left/ {\vphantom {1 {2\tau }}}
\right. \kern-\nulldelimiterspace} {2\tau }}$. The total number of
pulses to be stored simultaneously in the condensate of width $L$
for the duration
$
t_s = L/v_g
$
is found by
\begin{eqnarray}
C = \frac{L}{{2v_g \tau }},
\end{eqnarray}
which shows its limitation by the pulse spread. It is therefore
necessary to determine the actual bit storage capacity  by taking
into account higher order dispersive effects in slow light
propagation via EIT through atomic condensate.

\section{short pulse propagation through an atomic Bose-Einstein condensate\label{model}}

We calculated that for the present slow light experiments, third and higher
order dispersion effects are negligible. In fact, we have found the
third order dispersion coefficient is about seven orders of
magnitude smaller than the second order dispersion coefficient.
Taking into account this group velocity dispersion, the wave
equation describing slow short pulse propagation in the condensate can
be written by
\begin{eqnarray}
\frac{{\partial E}}{{\partial z}} + \alpha E + \frac{1}{{v_g
}}\frac{{\partial E}}{{\partial t}} + ib_2 \frac{{\partial ^2
E}}{{\partial t^2 }} = 0.
\end{eqnarray}
Here, the attenuation constant
\begin{eqnarray}
\alpha  =  - \frac{{i\pi }}{\lambda }\chi \left( {\omega _0 }
\right),
\end{eqnarray}
group velocity,
\begin{eqnarray}
\frac{1}{{v_g }} = \frac{1}{c} - \frac{\pi }{\lambda
}\frac{{\partial \chi }}{{\partial \omega }}\left( {\omega _0 }
\right),
\end{eqnarray}
and the second order dispersion coefficient,
\begin{eqnarray}
b_2  = \frac{\pi }{{2\lambda }}\frac{{\partial ^2 \chi
}}{{\partial \omega ^2 }}\left( {\omega _0 } \right)
\end{eqnarray}
are all calculated at the resonance. For the range $\sqrt
{4\pi c\left| {\mu _{31} } \right|^2 \rho/ \varepsilon
_0 \hbar}  \gg \Omega _c  \gg \Gamma _{2,3} $, they are given
by
\begin{eqnarray}
\alpha  &=& \frac{{2\pi \rho \left| {\mu _{31} } \right|^2 \Gamma
_2 }}{{\varepsilon _0 \hbar \lambda \Omega _c^2 }},\\
v_g  &=& \frac{{c\varepsilon _0 \hbar \Omega _c^2 }}{{2\omega _0
\left| {\mu _{31} } \right|^2 \rho }} \\
b_2  &=& i\frac{{8\pi \Gamma _3 \left| {\mu _{31} } \right|^2 \rho
}}{{\varepsilon _0 \hbar \lambda \Omega _c^4 }}.
\end{eqnarray}
The range of pump power can be established for $\Omega_c\sim
5\gamma - 15\gamma$ when ${\rho  \sim 10^{20}  - 10^{21}
{\rm{m}}^{ - 3} }$. It may be noted that broadening is more
sensitive to pump power than group velocity. This may be exploited
to optimize group delay and broadening for the purpose of maximum
bit storage capacity. For that aim let us now determine pulse
broadening and capacity analytically to see how they depend on
pump power. Analytical solution of the wave equation is possible
for a Gaussian pulse propagating through a uniform dispersive
medium. We consider an effective width $L$, given by the rms width of
the condensate, over which density can be considered uniform. This
treatment can be used for qualitative understanding of the pulse
shape variations in the condensate. After a delay time of
\begin{eqnarray}
t_d  = \frac{L}{{v_g }} - \frac{L}{c}
\end{eqnarray}
the pulse width will broaden into
\begin{eqnarray}
\tau \left( L \right) = \tau _0 \sqrt {1 + \left( {\frac{L}{{z_0
}}} \right)^2 }
\end{eqnarray}
where
\begin{eqnarray}
z_0  =  - \frac{{\pi \tau _0^2 }}{{b_2 }}.
\end{eqnarray}
When $L \gg z_0$ pulse spread can be determined from
\begin{eqnarray}
\tau \left( L \right) \cong \frac{{\left| {b_2 } \right|L}}{{\pi
\tau _0 }}.
\end{eqnarray}
Using $ C = L/2v_g \tau $ we find that
\begin{eqnarray}
C = \frac{L}{{2\tau _0 \sqrt {{{4\pi ^2 \Omega _c^4 }
\mathord{\left/
 {\vphantom {{4\pi ^2 \Omega _c^4 } {9\lambda ^4 \gamma ^2 \rho ^2  + {{4L^2 \Gamma _3^2 } \mathord{\left/
 {\vphantom {{4L^2 \Gamma _3^2 } {\pi ^2 \tau _0^4 \Omega _c^4 }}} \right.
 \kern-\nulldelimiterspace} {\pi ^2 \tau _0^4 \Omega _c^4 }}}}} \right.
 \kern-\nulldelimiterspace} {9\lambda ^4 \gamma ^2 \rho ^2  + {{4L^2 \Gamma _3^2 } \mathord{\left/
 {\vphantom {{4L^2 \Gamma _3^2 } {\pi ^2 \tau _0^4 \Omega _c^4 }}} \right.
 \kern-\nulldelimiterspace} {\pi ^2 \tau _0^4 \Omega _c^4 }}}}}
 }}.
\end{eqnarray}
We observe that two terms under the square root are competing with each other,
depending on the increase of the pump power.
At a critical Rabi frequency, given by
\begin{eqnarray}
\Omega _{c0}  = \left( {\frac{{3\Gamma _3 \gamma \lambda ^2 \left(
{\rho L} \right)}}{{\pi ^2 \tau _0^2 }}} \right)^{{1
\mathord{\left/
 {\vphantom {1 4}} \right.
 \kern-\nulldelimiterspace} 4}},
\end{eqnarray}
bit storage capacity becomes a maximum with a value
\begin{eqnarray}
C_{\max }  = \sqrt {\frac{{3\gamma \lambda ^2 \left( {\rho L}
\right)}}{{32\Gamma _3 }}}.
\end{eqnarray}
Storage time in this case turns out to be
\begin{eqnarray}
t_{s0}  = \frac{{\tau _0 \lambda }}{2}\sqrt {\frac{{3\gamma \left(
{\rho L} \right)}}{{\Gamma _3 }}}.
\end{eqnarray}
These expressions are all show dependence on density profile of
the cloud and using parameters of the trapping potential as well
as temperature and atomic scattering lengths, they can be tuned for
an optimum bit storage capacity obtained at a critical laser power.

Let us now examine the practical control parameters to optimize bir stroage capacity
by taking into account spatial inhomogeneity of cloud in
detail. To take it into account analytically, we consider a semi-ideal
BEC\cite{naraschewski98} where thermal and condensed
components overlap negligibly. This allows for separation of
atomic density in the form of
\begin{eqnarray}
\rho \left( {\vec r} \right) = \rho _c \left( {\vec r} \right) +
\rho _{th} \left( {\vec r} \right),
\end{eqnarray}
where
\begin{eqnarray}
\rho _c \left( {\vec r} \right) = \frac{{\mu  - V\left( {\vec r}
\right)}}{{u_0 }}\Theta \left( {\mu  - V\left( {\vec r} \right)}
\right)\Theta \left( {T_c  - T} \right)
\end{eqnarray}
represents the condensed component under Thomas-Fermi approximation
and
\begin{eqnarray}
\rho _{th} \left( {\vec r} \right) = \frac{{g_{{3 \mathord{\left/
 {\vphantom {3 2}} \right.
 \kern-\nulldelimiterspace} 2}}
\left[ {z{\rm{e}}^{ - \beta V\left( {\vec r} \right)} } \right]}}{{\lambda _T^3 }}
\end{eqnarray}
is the density of the thermal component of the cloud. Here
$
V\left( {\vec r} \right) = \left( {{m \mathord{\left/
 {\vphantom {m 2}} \right.
 \kern-\nulldelimiterspace} 2}} \right)\left( {\omega _r^2 r^2  + \omega _z^2 z^2 } \right)
$
is the trap potential written in cylindrical coordinates,
$ u_0 = {{4\pi \hbar ^2 a_s } \mathord{\left/
 {\vphantom {{4\pi \hbar ^2 a_s } m}} \right.
 \kern-\nulldelimiterspace} m}
$
is the atomic interaction coefficient. Fugacity is given by
$z = {\rm{e}}{}^{\beta \mu } $.
A Bose function is defined to be
\begin{eqnarray}
g_n \left( x \right) = \sum\limits_j {{{x^j } \mathord{\left/
 {\vphantom {{x^j } {j^n }}} \right.
 \kern-\nulldelimiterspace} {j^n }}}.
\end{eqnarray}
This model provides a good analytical fit to
the experimental BEC density profile at a finite temperature.
Chemical potential $\mu$ is found from $ N = \int {{\rm{d}}^3 \vec
r\rho \left( {\vec r} \right)}$. At low temperatures $T < T_c$
this yields
\begin{eqnarray}
\mu  = \mu _{TF} \left( {\frac{{N_0 }}{N}} \right)^{{2
\mathord{\left/
 {\vphantom {2 5}} \right.
 \kern-\nulldelimiterspace} 5}}
\end{eqnarray}
with
\begin{eqnarray}
\frac{{N_0 }}{N} = 1 - x^3  - s\frac{{\zeta \left( 2
\right)}}{{\zeta \left( 3 \right)}}x^2 \left( {1 - x^3 }
\right)^{{2 \mathord{\left/
 {\vphantom {2 5}} \right.
 \kern-\nulldelimiterspace} 5}}.
\end{eqnarray}
Here $x \equiv {T \mathord{\left/
 {\vphantom {T {T_c }}} \right.
 \kern-\nulldelimiterspace} {T_c }}
$ and $s$ is a scaling parameter. The semi-ideal model works best
for $s < 0.4$. From its definition
\begin{eqnarray}
s = \frac{{\mu _{{\rm{TF}}} }}{{k_{\rm{B}} T_{\rm{c}} }} =
\frac{1}{2}\zeta \left( 3 \right)^{{1 \mathord{\left/ {\vphantom
{1 3}} \right.
 \kern-\nulldelimiterspace} 3}} \left( {15N^{{1 \mathord{\left/
 {\vphantom {1 6}} \right.
 \kern-\nulldelimiterspace} 6}} \frac{{a_s }}{{a_h }}} \right)^{{2 \mathord{\left/
 {\vphantom {2 5}} \right.
 \kern-\nulldelimiterspace} 5}},
\end{eqnarray}
where $ a_h  = \sqrt {{\hbar  \mathord{\left/
 {\vphantom {\hbar  {m\left( {\omega _z \omega _r^2 } \right)^{{1 \mathord{\left/
 {\vphantom {1 3}} \right.
 \kern-\nulldelimiterspace} 3}} }}} \right.
 \kern-\nulldelimiterspace} {m\left( {\omega _z \omega _r^2 } \right)^{{1 \mathord{\left/
 {\vphantom {1 3}} \right.
 \kern-\nulldelimiterspace} 3}} }}},
$ we find that in slow light experiments semi-ideal model
description can be used up to scattering lengths $ a_s  <  \sim
8{\rm{ nm}}.$ It may be noted that at high temperatures
$T>T_{\rm{c}}$ chemical potential is determined from the usual
expression
\begin{eqnarray}
{\rm{Li}}_{\rm{3}} \left( z \right) = {{\zeta \left( 3 \right)}
\mathord{\left/
 {\vphantom {{\zeta \left( 3 \right)} {x^3 }}} \right.
 \kern-\nulldelimiterspace} {x^3 }},
\end{eqnarray}
where ${\rm{Li}}_{\rm{3}}$ is a polylogarithm. We also note
that at all temperatures it is required that $ \mu  \gg \hbar
\omega _{r,z}. $

Using the semi-ideal model, we can establish how propagation
parameters, $\alpha, v_g, b_2$ gain nonlocal character for
spatially nonuniform bosonic atomic cloud\cite{devrim06}. Assuming negligible spatial
inhomogeneity over the central region of width
\begin{eqnarray}
L = \left[ {\frac{{4\pi }}{N}\int\limits_0^\infty
{r{\rm{d}}r\int\limits_0^\infty  {{\rm{d}}zz^2 \rho \left( {r,z}
\right)} } } \right]^{{1 \mathord{\left/
 {\vphantom {1 2}} \right.
 \kern-\nulldelimiterspace} 2}}
\end{eqnarray}
and using the peak density value, which depends on temperature and
interatomic interactions as indicated by the semi-ideal model, we
can calculate the broadening of the pulse analytically\cite{devrim06}. As the
broadening is determined by the product $\rho L$, and $\rho$ and
$L$ exhibit competing dependencies on temperature, a maximum value
emerges just before the critical temperature in the broadening\cite{devrim06}.

Performing numerical simulations using Crank-Nicholson method
to solve the wave equation with spatially dependent coefficients,
we found that analytical method describes the behavior of
broadening very well qualitatively\cite{devrim06}, though quantitatively broadening is
overestimated by the analytical method. Bit storage capacity is found higher than
predicted by analytical calculation. The reason is that pulse
broadens as it approaches to more dispersive central region. As a
result it is guarded against further spreading. This is a unique
advantage of cold Bose gas in the light propagation due to its
characteristic density profile.

\section{Multiple optical modes within an atomic Bose-Einstein condensate\label{modes}}

In addition to diffraction\cite{ozgur}, effect of finite radial size of the cloud could be excitation
of multiple modes within the cloud.
We note that the refractive index for the radial
dimensions can be expressed as
\begin{eqnarray}
n(r) = \left\{ {\begin{array}{*{20}c}
   {n_0 \left[ {1 - 2\kappa \left( {\frac{r}{{R_{TF} }}} \right)^2 } \right]^{{1 \mathord{\left/
 {\vphantom {1 2}} \right.
 \kern-\nulldelimiterspace} 2}} ,} & {r \le R_{TF} ;}  \\
   {1,} & {r \ge R_{TF} .}  \\
\end{array}} \right.
\end{eqnarray}
at temperatures well below $T_C$. Here $ n_0  = \sqrt {1 + \chi
(0)}$ corresponds material refractive index at the center of the
cloud, $\kappa  = {{\chi (0)} \mathord{\left/
 {\vphantom {{\chi (0)} {4\left( {1 + \chi (0)} \right)}}} \right.
 \kern-\nulldelimiterspace} {4\left( {1 + \chi (0)} \right)}}
$, and $ R_{TF}  = \sqrt {{{2\mu (T)} \mathord{\left/
 {\vphantom {{2\mu (T)} {m\omega _r^2 }}} \right.
 \kern-\nulldelimiterspace} {m\omega _r^2 }}}
$ is the radial Thomas-Fermi radius. It should be noted that under EIT conditions,
thermal cloud and the condensed component would have same refractive index at EIT
resonance. On the other hand, slightly detuned pulse may excite multiple modes in the
condensate.
We now establish conditions for single mode and multi-mode excitations for such a case.

Ignoring the group velocity
dispersion in the radial dimensions, the wave equation $ \left(
{\nabla ^2  + k^2 } \right)E_t  = 0 $, with $ k = {{\omega n}
\mathord{\left/
 {\vphantom {{\omega n} c}} \right.
 \kern-\nulldelimiterspace} c}
$
 can be solved using an ansatz of the form $
E_t  = \psi (r){\rm{e}}^{il\phi } {\rm{e}}^{i\left( {\omega t -
\beta z} \right)}. $ This yields
\begin{eqnarray}
\left[ {\frac{{d^2 }}{{dr^2 }} + \frac{1}{r}\frac{d}{{dr}} + p^2
(r)} \right]\psi (r) = 0
\end{eqnarray}
where
\begin{eqnarray}
p^2 (r) = k_0^2 n^2 (r) - \beta ^2  - \frac{{l^2 }}{{r^2 }}
\end{eqnarray}
with
$
k_0  = {\omega  \mathord{\left/
 {\vphantom {\omega  c}} \right.
 \kern-\nulldelimiterspace} c}.
$ Number of modes supported by the condensate can be estimated by
a simple WKB treatment. Quantization condition
\begin{eqnarray}
\int_{r_1 }^{r_2 } {p(r){\rm{d}}r = \left( {m + \frac{1}{2}}
\right)\pi } {\rm{   }}m = 0,1,2,...
\end{eqnarray}
leads to the propagation constant
\begin{eqnarray}
\beta _{lm}  = n{}_0k_0 \left[ {1 - \frac{{2\sqrt {2\kappa }
}}{{n{}_0k_0 R_{TF} }}\left( {l + 2m + 1} \right)} \right]^{{1
\mathord{\left/
 {\vphantom {1 2}} \right.
 \kern-\nulldelimiterspace} 2}},
\end{eqnarray}
so that the number of modes are determined from the constraint
\begin{eqnarray}
2(l + 2m + 1) \le n_0 k_0 R_{TF} \sqrt {2\kappa }.
\end{eqnarray}
Single mode condition for a BEC can be imposed in terms of radial
size of the atomic cloud. Using experimental parameters we find that single mode
condition at $T=42\,$nK can be expressed in terms of the radial
radius of the cloud being $R<1.13\,\mu$m. Around $T_c$ this radius
can be about $\sim 3\,\mu$m.

Multimode propagation could be useful for handling many
transmission channels on one material. Short axial size of BEC and
its unique density profile can be exploited to keep modal
dispersion weak. In such short distance applications, this may be
utilized to capture optimum amount of power from a light source
and enhance the optical coupling capability of BEC. The
modes closer to the thermal envelope, where the cloud is less
dense, have faster group velocities than those in the central region. This results in
pulse spread less than the case of a more uniform cloud.
Because of Thomas-Fermi density profile, modal dispersion in the condensate becomes more significant at
lower temperatures. Finally we propose that more index contrast and hence support of more
modes in the BEC can be possible by some other quantum coherent effects such as index enhancement,
alternative to off-resonant EIT, if one demands multiple mode excitations in the BEC.

\section{Conclusion \label{summary}}
Suumarizing, we have studied short pulse propagation under EIT conditions within an atomic semi-ideal
BEC. Third and higher order dispersive effects were found to be negligible in
current BEC systems for propagation of pulses of widths up to
nanoseconds. For pulses of widths greater than microseconds second
order dispersion was also found to be insignificant. We have found that just below the critical
temperature, broadening of the pulse becomes maximum. Besides, for a semi-ideal
BEC, we show that broadening decreases with increasing scattering length. Taking into account
spatial inhomegenity of the cloud, we conclude that
axial spatial variation  of the cloud guards the pulse against spreading due to
group velocity dispersion. As the pulse broadens while it
gets closer and closer to central region, it suffers less and less dispersive
effects. Analytical studies indicated that there is a critical pump power which gives maximum
achievable bit storage capacity. This value was shown to be independent of
initial pulse width, but strongly depends on density profile of the
atomic cloud. Multimode pulse propagation due to finite radial size was also examined. Single
mode condition for light propagation was established for atomic BEC. Either
off-resonant EIT or quantum coherent index enhancement can be used
to realize and enhance multi-mode excitations. We conclude that radial density profile of BEC protects the
pulse against spreading due to modal dispersion.
\acknowledgements

We would like to thank N. Postac\i{}o\~{g}lu and A. Sennaroglu for
valuable discussions. \"O.E.M. gratefully
acknowledges financial support from the Young Scientist Program
(GEBIP) of the Turkish Academy of Sciences (T\"UBA). This work was
partially supported by Istanbul Technical University Foundation
(ITU-BAP) under Project No. 31192.


\newpage

Corresponding authors:\\
\"O. E. M\"ustecapl{\i}o\u{g}lu\\
Office telephone number: +90-(212)-338 1424\\
Home telephone number: +90-(212)-338 3149\\
Fax number: +90-(212)-338 1559\\
E-Mail address: omustecap@ku.edu.tr\\
{\bf Devrim Tarhan} \\
E-Mail address:devrimtarhan@yahoo.com,dtarhan@gmail.com


\begin{thebibliography}{}

\bibitem{hau99} L.V. Hau, S.E. Harris, Z. Dutton, and C.H. Behroozi,
 \nat {\bf 397,} 594 (1999).

\bibitem{inouye00} S. Inouye, R.F.Löw, S. Gupta, T. Pfau, A. Görlitz, T. L. Gustavson, D. E. Pritchard
and W. Ketterle, \prl {\bf 85,} 4225-4228 (2000).

\bibitem{kash99} M. M. Kash, V. A. Sautenkov, A. S. Zibrov, L. Hollberg,
G. R. Welch, M. D. Lukin, Y. Rostovtsev, E. S. Fry, and M. O.
Scully,  \prl {\bf 82,} 5229 (1999).

\bibitem{budker99} D. Budker, D. F. Kimball, S. M. Rochester, and V. V. Yashchuk,
\prl {\bf 83,} 1767 (1999).

\bibitem{lukin00} M. D. Lukin and A. Imamoglu,
\prl {\bf 84,} 1419 (2000).

\bibitem{lukin01} M. D. Lukin and A. Imamoðlu,
\nat {\bf 413,}, 273 (2001).

\bibitem{matsko00} A. B. Matsko, Y. V. Rostovtsev, H. Z. Cummins, and M. O. Scully,
\prl {\bf 84,} 5752 (2000).

\bibitem{liu01} C. Liu, Z. Dutton, C. H. Behroozi, and L. V. Hau,
\nat {\bf 409,} 490 (2001).

\bibitem{agarwal02} G.S. Agarwal and S. Das Gupta, Phy. Rev. A 65, 053811, 1-5, 2002.

\bibitem{wei05} X.-G. Wei, J.-H. Wu, G.-X. Sun, Z. Shao, Z.-H. Kang, Y. Jiang, and J.-Y. Gao,
\pra {\bf 72,} 023806 (2005).

\bibitem{sautenkov05} V. A. Sautenkov, Y. V. Rostovtsev, C. Y. Ye, G. R. Welch,
O. Kocharovskaya, and M. O. Scully, \pra {\bf 71,} 063804 (2005).

\bibitem{sun05} Q. Sun, Y. V. Rostovtsev, J. P. Dowling, M. O. Scully, and M. S. Zubairy,
\pra {\bf 72,} 031802(R) (2005).

\bibitem{bashkansky05} M. Bashkansky, G. Beadie, Z. Dutton, F. K. Fatemi, J. Reintjes,
and M. Steiner, \pra {\bf 72,} 033819 (2005).

\bibitem{naraschewski98} M. Naraschewski, D. M. Stamper-Kurn,
\pra {\bf58,} 2423 (1998).

\bibitem{devrim06} D. Tarhan, A. Sennaroglu, \"O. E. M\"ustecapl{\i}o\u{g}lu,
J. Opt. Soc. Am. B , {\bf 23(9),} 1925-1933 (2006).

\bibitem{ozgur} \"O. E. M\"ustecapl{\i}o\u{g}lu and L. You, Opt. Commun. {\bf 193,}
301-312 (2001).

\end{thebibliography}
\end{document}